\documentclass[conference,a4paper]{IEEEtran}

\usepackage[utf8]{inputenc} 
\usepackage[T1]{fontenc}
\usepackage{url}              % provides \url{...}
\usepackage{cite}             % improves presentation of citations

\usepackage[cmex10]{amsmath}  % Use the [cmex10] option to ensure complicance
\usepackage{mleftright}       % fix to wrong spacing of \left-,
\mleftright                   % \middle- \right-commands 

\usepackage{graphicx}         % provides \includegraphics{...} to
\usepackage{booktabs}         % fixes poor spacing in tables and
\newtheorem{lem}{Lemma}
\newtheorem{coro}{Corollary}

\newtheorem{thm}{Theorem}

\newtheorem{remark}{Remark}

\usepackage{amsfonts}

\begin{document}

\title{CFMA for Gaussian MIMO Multiple Access Channels}

%%%%%%
% \author{%
%   \IEEEauthorblockN{Anonymous Authors}
%   % \IEEEauthorblockA{%
%   %   Please do NOT provide authors' names and affiliations\\
%   %   in the paper submitted for review, but keep this placeholder.\\
%   %   ISIT23 follows a \textbf{double-blind reviewing policy}.}
% }

%%%%%% Please only add the author names and affiliations for the FINAL
%%%%%% version of the paper, but NOT for the paper submitted for review!
%
%%%%%
%%%%% Single author, or several authors with same affiliation:
\author{%
  \IEEEauthorblockN{Lanwei Zhang, Jamie Evans and Jingge Zhu}
  \IEEEauthorblockA{The University of Melbourne\\
                    Victoria 3010 Australia\\
                    lanweiz@student.unimelb.edu.au}
                  }

\maketitle

%%%%%
%% Abstract: 
%% If your paper is eligible for the student paper award, please add
%% the comment "THIS PAPER IS ELIGIBLE FOR THE STUDENT PAPER
%% AWARD." as a first line in the abstract. 
%% For the final version of the accepted paper, please do not forget
%% to remove this comment!
%%
% \begin{abstract}
%   THIS PAPER IS ELIGIBLE FOR THE STUDENT PAPER AWARD.
%   (Remove this comment for final version of paper.)
%   Instructions are given for the preparation and submission of papers
%   for the \emph{2023 International Symposium on Information
%     Theory}. This template is available (including \LaTeX{}-source)
%   from \url{https://isit2023.org/}
% \end{abstract}
\begin{abstract}
Compute-forward multiple access (CFMA) is a multiple access transmission scheme based on Compute-and-Forward (CF) which allows the receiver to first decode linear combinations of the transmitted signals and then solve for individual messages. This paper extends the CFMA scheme to a two-user Gaussian multiple-input multiple-output (MIMO) multiple access channel (MAC). We first derive the expression of the achievable rate pair for MIMO MAC with CFMA. We prove a general condition under which CFMA can achieve the sum capacity of the channel. Furthermore, this result is specialized to SIMO and $2$-by-$2$ diagonal MIMO multiple access channels, for which more explicit sum capacity-achieving conditions on power and channel matrices are derived. Numerical results are also provided for the performance of CFMA on general MIMO multiple access channels.
\end{abstract}

\vspace{-1mm}
\section{Introduction}\label{sec_intro}
In a wireless network, interference from other transmitters can be problematic for a receiver to recover the desired message. The traditional orthogonal transmission schemes avoid interference by allocating the transmission resources orthogonally to every transmitter. For example, the time-division multiple access (TDMA) divides each transmitted signal into different time slots. However, these methods suffer from a diminishing rate, especially when many transmitters send messages simultaneously \cite{Tse05fundamentalsof}.

Compute-and-forward (CF) is a linear physical-layer network coding scheme proposed in \cite{Nazer11CF}, which allows the receiver to decode a linear combination of the messages from multiple transmitters. Then the receiver can either forward the linear combination to the next receiver, thus working as a relay, or collect enough linear combinations to recover each message. In this way, it does not avoid interference but exploits it in the decoding process. The key to the CF scheme is the adoption of nested lattice codes, which guarantee that any integer linear combination of the codewords is still a codeword. Lattice coding and decoding have been used in MIMO channels and show good performance. The class of LAttice Space–Time (LAST) codes has been proposed in \cite{Gamal04} and achieves the optimal diversity–multiplexing tradeoff of MIMO Channels. In \cite{Ordentlich15}, the authors show that with a precoded integer-forcing scheme, which is also based on the nested lattice codes, they can achieve the optimal MIMO capacity to within a constant gap. 

Compute-forward multiple access (CFMA) is a generalized CF scheme proposed in \cite{Zhu17}, which allows the users to have different rates while the rates are the same for all the users in the original CF. It shows that in a two-user single-input single-output (SISO) MAC, where each transmitter or the receiver only has a single antenna, CFMA can achieve the whole capacity pentagon without time sharing. Since multiple linear combinations are required to recover each transmitted signal, successive cancellation decoding is applied, where the previous decoded linear combinations are used in the later decoding process. 

This paper extends the CFMA scheme to a two-user Gaussian MIMO MAC, where the transmitters and the receiver have multiple antennas. A concatenating channel model is applied, where all the transmission symbols are concatenated into a vector. We first give the analytical expressions of the achievable rate pair. They are the generalized version of the SISO case in \cite{Zhu17}. Then we show that the MIMO MAC sum capacity is achievable given certain polynomial-form conditions. Although the conditions are not always satisfiable for all power constraints and channel conditions, we provide some capacity-achieving guarantees when the power constraint is large enough for some special cases, including single-input multiple-output (SIMO) MAC and diagonal MIMO MAC. We note that in \cite{Nazer16}, the authors proposed an expanded CF scheme for SIMO MAC and show that their scheme can achieve the sum capacity given some power constraints and channel conditions. However, their results do not take an explicit form. As one of the contributions in this paper, we give more concise sum capacity-achieving conditions for the two-user SIMO channels.

\vspace{-1mm}
\section{System Model}\label{sec_model}
We consider a two-user Gaussian MIMO MAC where each user is equipped with $t$ transmit antennas and the receiver is equipped with $r$ receive antennas. The channel coefficients are assumed stationary and known to both transmitters and receivers during the transmissions. They are denoted as $\textbf{H}_l$ with dimension $r\times t$ for $l=1,2$. The channel input for each user is constrained by a given covariance matrix $\textbf{K}_l\succeq \textbf{0}$ with dimension $t\times t$, which satisfies the power constraint $tr(\textbf{K}_l) \leq P$ for $l=1,2$, where $tr(\cdot)$ refers to the trace of a matrix. We use $\textbf{x}_{l,s}\in \mathbb R^t$ to denote the channel input of user $l$ at time $s$. At time $s$, the channel output is given by
\begin{equation}
    \label{eq_channel_output_MIMO_single_channel_use}
    \textbf{y}_s = \sum_{l=1}^2 \textbf{H}_l \textbf{x}_{l,s} + \textbf{z}_s,
\end{equation}
where $\textbf{z}_s$ is the additive white Gaussian noise with zero mean and unit variance. Let $n$ denote the total number of channel uses (hence the length of the codebook). To simplify the presentation, we apply a concatenating channel model \cite{Gamal04}\cite{Lin11}, where all the $n$ symbols are concatenated into a $tn$-dimensional vector, i.e., $\textbf{x}_l = (\textbf{x}_{l,1}^T,\ldots,\textbf{x}_{l,n}^T)^T$. In this case, the channel output for all the channel uses is given by
\begin{equation}
    \label{eq_channel_output_MIMO_vector}
    \textbf{y} = \sum_{l=1}^2 \bar{\textbf{H}}_l \textbf{x}_l + \textbf{z},
\end{equation}
where $\textbf{y}$ and $\textbf{z}$ are also concatenating vectors with dimension $rn$, and $\bar{\textbf{H}}_l := \textbf{I}_n \otimes \textbf{H}_l$ are $rn \times tn$ block diagonal matrices. Note that $\textbf{I}_n$ is the $n$-dimensional identity matrix, and $\otimes$ denotes the Kronecker product. The channel output covariance matrix of each user in this concatenating model is given by $\bar{\textbf{K}}_l := \textbf{I}_n \otimes \textbf{K}_l$, for $l = 1,2$.

\section{Coding scheme}\label{sec_coding}
In this section, we extend the CFMA scheme in \cite{Zhu17} from SISO MAC to the two-user MIMO MAC. The codebook of user $l$ is constructed by a $tn$-dimensional nested lattice denoted by
\begin{equation}\label{eq_codebook_MIMO}
    \mathcal{C}_l = \Lambda_l^F\cap \mathcal{V}_l^C,
\end{equation}
where the fine lattice $\Lambda_l^F$ is chosen to be good for AWGN channel coding in the sense of \cite{Erez05}, and $\mathcal{V}_l^C$ is the Voronoi region of the coarse lattice $\Lambda_l^C$, which is chosen to be good for quantization in the sense of \cite{Erez05}. We denote the second moment of the coarse lattice as
\begin{equation}\label{eq_second_moment_MIMO}
    \frac{1}{tn\text{Vol}(\mathcal{V}_l^C)}\int_{\mathcal{V}_l^C} ||\textbf{x}||^2d\textbf{x} = \beta_l^2,\quad\beta_l\in\mathbb{R}.
\end{equation}
With this codebook, the message rate of user $l$ is given by
\begin{equation}\label{eq_message_rate_MIMO}
    r_l = \frac{\log|\mathcal{C}_l|}{n} = \frac{1}{n}\log\frac{\text{Vol}(\mathcal{V}_l^C)}{\text{Vol}(\mathcal{V}_l^F)}.
\end{equation}

Given a codeword $\textbf{t}_l$, we can define $\textbf{c}_l = [\textbf{t}_l/\beta_l+\textbf{d}_l]\mod \Lambda_l^C /\beta_l$, where $\textbf{d}_l$ is the dither vector which is uniformly distributed in the region $\Lambda_l^C /\beta_l$, and $\mod$ is the modulo operation. We can also write $\textbf{c}_l$ as
\begin{equation}
    \label{eq_c_l}
    \textbf{c}_l = (\textbf{t}_l/\beta_l+\textbf{d}_l) - \mathcal{Q}_{\Lambda_l^C /\beta_l}(\textbf{t}_l/\beta_l+\textbf{d}_l),
\end{equation}
where $\mathcal{Q}_\Lambda$ is the quantization operation over the lattice $\Lambda$. Readers are referred to \cite{Nazer11CF,Zhu17} for more details about the nested lattice codes. The channel input is generated as
\begin{equation}
    \label{eq_channel_input_MIMO}
    \textbf{x}_l = \bar{\textbf{B}}_l  \textbf{c}_l,
\end{equation}
where $\bar{\textbf{B}}_l: = \textbf{I}_n \otimes \textbf{B}_l$ is a $tn\times tn$ lower triangular matrix such that $\textbf{K}_l = \textbf{B}_l \textbf{B}_l^T$. Since $\textbf{K}_l = \mathbb{E} [\textbf{x}_l\textbf{x}_l^T]$ is always positive semi-definite, we can get the unique $\textbf{B}_l$ by Cholesky decomposition, which is a $t\times t$ lower triangular matrix with non-negative diagonal. 

In the CFMA scheme, the receiver will decode integer linear combinations of the lattice codewords instead of the individual symbols. A typical form is given by
\begin{equation}
    \label{eq_decode_function_MIMO}
    \textbf{u}:=\left[\sum_{l=1}^2 a_l \textbf{t}_l \right]\mod \Lambda_f^C,
\end{equation}
where $a_l\in\mathbb{Z}$, and $\Lambda_f^C$ denotes the finest lattice among $\Lambda_l^C$. In order to recover the individual messages, the receiver can decode two linearly independent integer combinations and solve for the individual codewords. In addition, after decoding the first linear combination, the receiver can use it in the decoding process for the second linear combination, i.e., successive cancellation decoding. In the next section, we will discuss how the two linear combinations are computed and derive the expressions of achievable rate pairs.

\section{Achievable Rates}\label{sec_rate}
\begin{thm}\label{thm_rate_pair}
For a two-user MIMO Gaussian MAC given the channel matrices $\textbf{H}_1,\textbf{H}_2$ and the input covariance matrices $\textbf{K}_1,\textbf{K}_2$, with the CFMA scheme, the following rate pair is achievable
\begin{equation}\label{eq_achievable_rate_MIMO_MAC_CFMA}
    R_l = \left\{\begin{array}{ll}
        r_l(\textbf{a},\boldsymbol{\beta}), & b_l = 0 \\
        r_l(\textbf{b}|\textbf{a},\boldsymbol{\beta}), & a_l = 0 \\
        \min\{r_l(\textbf{a},\boldsymbol{\beta}),\;r_l(\textbf{b}|\textbf{a},\boldsymbol{\beta})\}, & \text{otherwise}
    \end{array}\right.
\end{equation}
for any linearly independent $\textbf{a},\textbf{b}\in\mathbb{Z}^2$ and $\boldsymbol{\beta}\in\mathbb{R}^2$ if $r_l(\textbf{a},\boldsymbol{\beta})\geq 0$ and $r_l(\textbf{b}|\textbf{a},\boldsymbol{\beta}) \geq 0$ for $l = 1,2$, where
\begin{align}
    r_l(\textbf{a},\boldsymbol{\beta}) = & \frac{1}{2} \log 
    \frac{\beta_l^{2t} \left|\textbf{I}_{r} + \textbf{H}_1 \textbf{K}_1\textbf{H}_1^T + \textbf{H}_2 \textbf{K}_2\textbf{H}_2^T\right|}
    {\left| \textbf{M} \right|} \label{eq_achievable_rate_MIMO_first}\\
    r_l(\textbf{b}|\textbf{a},\boldsymbol{\beta}) = & \frac{1}{2} \log
    \frac{\beta_l^{2t} {\left| \textbf{M} \right|}}
    {(\tilde a_1 \tilde b_2 - \tilde a_2 \tilde b_1)^{2t}} \label{eq_achievable_rate_MIMO_second}
\end{align}
with $\tilde a_l = a_l \beta_l, \tilde b_l = b_l \beta_l$ and 
\begin{equation}
    \label{eq_Mh}
    \textbf{M} = (\tilde a_1^2 + \tilde a_2^2)\textbf{I}_{t} + (\tilde a_1 \textbf{B}_2^T \textbf{H}_2^T - \tilde a_2 \textbf{B}_1^T \textbf{H}_1^T)(\tilde a_1 \textbf{H}_2 \textbf{B}_2 - \tilde a_2 \textbf{H}_1 \textbf{B}_1).
\end{equation}
\end{thm}

\begin{IEEEproof}
The computation coefficients of the first linear combination are denoted as $\textbf{a} = (a_1,a_2)^T$. Let the receiver compute
\begin{align*}
    \textbf{y}_1^\prime = & \textbf{W} \textbf{y} -\sum_l a_l\beta_l \textbf{d}_l\\
    = & \textbf{W}(\sum_l\bar{\textbf{H}}_l \bar{\textbf{B}}_l  \textbf{c}_l + \textbf{z})-\sum_l a_l\beta_l \textbf{d}_l -\sum_l a_l\beta_l \textbf{c}_l+\underbrace{\sum_l a_l\beta_l \textbf{c}_l}_{\text{use } (\ref{eq_c_l})}\\
    = & \underbrace{\textbf{W}\textbf{z} + \sum_l (\tilde a_l\textbf{I}_{tn}-\textbf{W}\bar{\textbf{H}}_l\bar{\textbf{B}}_l)(-\textbf{c}_l)}_{\bar{\textbf{z}}_1}-\sum_l a_l\beta_l \textbf{d}_l\\
    & + \sum_l a_l\beta_l(\textbf{t}_l/\beta_l+\textbf{d}_l) - a_l\beta_l\mathcal{Q}_{\Lambda_l^C /\beta_l}(\textbf{t}_l/\beta_l+\textbf{d}_l) \\
    = & \bar{\textbf{z}}_1 + \sum_l a_l \underbrace{\left(\textbf{t}_l - \mathcal{Q}_{\Lambda_l^C }(\textbf{t}_l+\beta_l\textbf{d}_l)\right)}_{\tilde{\textbf{t}}_l},
\end{align*}
where $\textbf{W}$ is a $tn\times rn$ equalization matrix, which will be determined later. The last equality holds because $\mathcal{Q}_\Lambda(\beta \textbf{x}) = \beta \mathcal{Q}_{\Lambda/\beta}(\textbf{x})$ for any lattice $\Lambda$ and any nonzero real $\beta$. It is enough to recover the message from $\tilde{\textbf{t}}_l$ since $\tilde{\textbf{t}}_l$ and the codeword $\textbf{t}_l$ belong to the same coset of $\Lambda_f^C$ \cite{Zhu17,Gamal04}. With lattice decoding, the achievable rate to decode $\sum_l a_l \tilde{\textbf{t}}_l$ is given by
\begin{equation}\label{eq_achievable_rate_general}
    r_{l}^{(1)} = \max_{\textbf{W}}\quad \frac{1}{2n} \log^+ \frac{|\beta_l^2\textbf{I}_{tn}|}{|\boldsymbol{\Sigma}_1(\textbf{W})|},
\end{equation}
where $|\cdot|$ refers to the determinant, and $\boldsymbol{\Sigma}_1(\textbf{W}) = \textbf{W}\textbf{W}^T + \sum_l (\tilde a_l\textbf{I}_{tn}-\textbf{W}\bar{\textbf{H}}_l\bar{\textbf{B}}_l)(\tilde a_l\textbf{I}_{tn}-\textbf{W}\bar{\textbf{H}}_l\bar{\textbf{B}}_l)^T$. For brevity, we may use $\boldsymbol{\Sigma}_1$ to refer to $\boldsymbol{\Sigma}_1(\textbf{W})$. Note that $\boldsymbol{\Sigma}_1$ is positive definite. To prove the achievable rate in (\ref{eq_achievable_rate_general}), we will apply the ambiguity lattice decoder of \cite{Loeliger97} with the decision region
\begin{equation}
    \label{eq_decision_region}
    \mathcal{E}_{n,\eta} = \{\textbf{z}\in\mathbb{R}^{tn}: |\textbf{Q}\textbf{z}|^2\leq tn(1+\eta)\},
\end{equation}
where $\eta>0$ and $\textbf{Q}\in\mathbb{R}^{tn\times tn}$ satisfies $\textbf{Q}^T\textbf{Q} = \boldsymbol{\Sigma}_1^{-1}$. Recall that we are considering an ensemble of $tn$-dimensional nested lattices $\{\Lambda_l^C \subseteq \Lambda_l^F\}$ for $l=1,2$ with message rate given by (\ref{eq_message_rate_MIMO}). Now we let $\Lambda_C = \Lambda_l^C / \beta_l$ with Voronoi region $\mathcal{V}_C$ thus $\sigma^2(\Lambda_C) = 1$. We further let $\Lambda_F = \Lambda_l^F$ for $l=1,2$, and its fundamental volume $\text{Vol}(\mathcal{V}_F)$ is fixed and constant with $n$ where $\mathcal{V}_F$ is the Voronoi region of $\Lambda_F$. For the equivalent channel $\textbf{y}_1^\prime = \bar{\textbf{z}}_1 + \textbf{u}$ where $\textbf{u} \in \Lambda_F$, the error probability is upper-bounded by
\begin{equation}
    \label{eq_error_prob}
    P_e(\mathcal{E}_{n,\eta}|\textbf{u}) \leq Pr(\hat{\textbf{u}} \neq \textbf{u}) + Pr(\mathcal{A}),
\end{equation}
where $\hat{\textbf{u}}$ is the decoded codeword and $\mathcal{A}$ is the ambiguity event defined as the event that the received point $\textbf{y}_1^\prime$ belongs to $\{\mathcal{E}_{n,\eta} + \textbf{u}_1\} \cap \{\mathcal{E}_{n,\eta} + \textbf{u}_2\}$ for some pair of distinct lattice points $\textbf{u}_1,\textbf{u}_2\in \Lambda_F$. By taking the expectation over the ensemble of random
lattices, from \cite[Theorem 4]{Loeliger97} we obtain the upper bound of the average error probability 
\begin{equation}
    \label{eq_average_error_prob}
    \bar P_e(\mathcal{E}_{n,\eta}) \leq Pr(\bar{\textbf{z}}_1 \notin \mathcal{E}_{n,\eta}) + (1+\delta)\frac{\text{Vol}(\mathcal{E}_{n,\eta})}{\text{Vol}(\mathcal{V}_F)},
\end{equation}
where $\delta>0$ and the term $\text{Vol}(\mathcal{E}_{n,\eta})$ refers to the volume of the decision region which can be represented by
\begin{equation}
    \label{eq_vol_decision_region}
    \text{Vol}(\mathcal{E}_{n,\eta}) = (1+\eta)^{tn/2}|\textbf{Q}^T\textbf{Q}|^{-1/2}\text{Vol}(\mathcal{B}(\sqrt{tn})),
\end{equation}
where $\text{Vol}(\mathcal{B}(\sqrt{tn}))$ is the volume of a $tn$-dimensional sphere of radius $\sqrt{tn}$. For brevity, we will still use $Pr(\mathcal{A})$ to represent the second term of (\ref{eq_average_error_prob}). From (\ref{eq_message_rate_MIMO}), we know 
\begin{equation}
    \label{eq_vol_F}
    \text{Vol}(\mathcal{V}_F) = \text{Vol}(\mathcal{V}_l^C)\cdot2^{-n r_l} = \beta_l^{tn}\text{Vol}(\mathcal{V}_C)\cdot2^{-n r_l}.
\end{equation}
By combining (\ref{eq_vol_decision_region}) and (\ref{eq_vol_F}), we have
\begin{equation}
    \label{eq_ambiguity_prob_rate}
    Pr(\mathcal{A}) \leq (1+\delta)\cdot 2^{-n\left[\frac{1}{2n}\log\frac{\beta_l^{2tn}}{|\boldsymbol{\Sigma}_1|}-r_l-\eta'\right]}
\end{equation}
where $\eta' = \frac{t}{2}\log(1+\eta)+\frac{1}{n}\log \frac{\text{Vol}(\mathcal{B}(\sqrt{tn}))}{\text{Vol}(\mathcal{V}_C)}$. As $n\rightarrow\infty$, we have $G(\Lambda_C)\rightarrow \frac{1}{2\pi e}$ by the goodness for covering, where $G(\Lambda_C)$ is the normalized second-order moment of $\Lambda_C$. Given $\text{Vol}(\mathcal{V}_C) = \left(\frac{\sigma^2(\Lambda_C)}{G(\Lambda_C)}\right)^{tn/2}$, we have $\text{Vol}(\mathcal{V}_C) \rightarrow (2\pi e)^{tn/2}$ when $n\rightarrow\infty$.
By using the fact
\begin{equation}
    \label{eq_vol_ball_inf}
    \text{Vol}(\mathcal{B}(\sqrt{tn})) \rightarrow \frac{(2\pi e)^ {tn/2}}{\sqrt{tn\pi}} \text{ as } n\rightarrow\infty, 
\end{equation}
we have
\begin{equation}
    \label{eq_vol_fraction_inf}
    \frac{1}{n}\log \frac{\text{Vol}(\mathcal{B}(\sqrt{tn}))}{\text{Vol}(\mathcal{V}_C)} \rightarrow -\log (\sqrt{tn\pi})^{1/n} \rightarrow 0.
\end{equation}
The second right arrow comes from the fact that $\lim_{x\rightarrow\infty}(x)^{\lambda/x}=1,\forall\lambda>0$ when $x = tn\pi$ and $\lambda = \frac{t\pi}{2}$. Since $\eta>0$ is arbitrary, we have $\eta'\rightarrow 0$ as $n\rightarrow\infty$. Therefore, we can conclude that for arbitrary $\epsilon_1 >0$, $Pr(\mathcal{A})\leq \epsilon_1/2$ for sufficiently large $n$ given that 
\begin{equation}
    \label{eq_rate_ambiguity}
    r_l < \frac{1}{2n}\log\frac{\beta_l^{2tn}}{|\boldsymbol{\Sigma}_1|}.
\end{equation}
The proof will then be complete if we can show that $Pr(\bar{\textbf{z}}_1 \notin \mathcal{E}_{n,\eta})\leq \epsilon_2/2$ for arbitrary $\eta, \epsilon_2>0$. Recall that 
\begin{equation}
    \label{eq_wrong_decode_error}
    Pr(\bar{\textbf{z}}_1 \notin \mathcal{E}_{n,\eta}) = Pr(|\textbf{Q}\bar{\textbf{z}}_1|^2 > tn(1+\eta)),
\end{equation}
where $\bar{\textbf{z}}_1 = \textbf{W}\textbf{z} + \sum_l (\tilde a_l\textbf{I}_{tn}-\textbf{W}\bar{\textbf{H}}_l\bar{\textbf{B}}_l)(-\textbf{c}_l)$ is the effective noise with $\textbf{z}\sim \mathcal{N}(\textbf{0},\textbf{I}_{rn})$ and $\textbf{c}_l\sim \text{Uniform}(\mathcal{V}_C)$ for $l=1,2$. Note that $\textbf{z}, \textbf{c}_1, \textbf{c}_2$ are statistically independent. To upper-bound this probability, we will consider a ``noisier" system with higher noise variance. We first add a Gaussian vector $\textbf{e}_3\sim \mathcal{N}(\textbf{0},(\sigma^2-1)\textbf{I}_{rn})$ to $\textbf{z}$ to make it noisier, where 
\begin{equation}
    \sigma^2 = \frac{r_{c}(\Lambda_C)^2}{tn}.
\end{equation}
The term $r_{c}(\Lambda_C)$ is the covering radius of $\Lambda_C$. From \cite{Gamal04}, we know $\sigma^2>1$. As $\sigma^2-1>0$, the noise $\textbf{e}_3$ is well-defined. We then replace $\textbf{c}_l$ with Gaussian vector $\textbf{e}_l\sim\mathcal{N}(\textbf{0},\sigma^2\textbf{I}_{tn})$. Recall $\sigma^2(\Lambda_C) =1$, thus $\textbf{e}_l$ has a larger variance. The considered noise is then replaced by
\begin{equation}
    \label{eq_noisier_noise}
    \bar{\textbf{z}}_1' = \textbf{W}(\textbf{z}+\textbf{e}_3) + \sum_l (\tilde a_l\textbf{I}_{tn}-\textbf{W}\bar{\textbf{H}}_l\bar{\textbf{B}}_l)\textbf{e}_l.
\end{equation}
Let $f_{\textbf{c}_l}(\cdot)$ and $f_{\textbf{e}_l}(\cdot)$ denote the PDF of $\textbf{c}_l$ 
 and $\textbf{e}_l$, respectively. Following the argument of \cite[Lemma 11]{Erez04} and \cite{Gamal04}, we obtain
\begin{equation}
    \label{eq_pdf_upper_bound}
    f_{\textbf{c}_l}(\textbf{z}) \leq \left(\frac{r_{c}(\Lambda_C)}{r_{e}(\Lambda_C)}\right)^{tn} \exp(o(tn)) f_{\textbf{e}_l}(\textbf{z}),
\end{equation}
where $r_{e}(\Lambda_C)$ is the effective radius of $\Lambda_C$. When $n\rightarrow\infty$, $r_{c}(\Lambda_C)/r_{e}(\Lambda_C)\rightarrow 1$ from the goodness for covering. From the noisier construction, (\ref{eq_wrong_decode_error}) can be upper-bounded by
\begin{equation}
    \label{eq_wrong_decode_error_upper_bound}
    \begin{split}
        &Pr(|\textbf{Q}\bar{\textbf{z}}_1|^2 > tn(1+\eta)) \\ &\leq \left[\left(\frac{r_{c}(\Lambda_C)}{r_{e}(\Lambda_C)}\right)^{tn} \exp(o(tn))\right]^2 Pr(|\textbf{Q}\bar{\textbf{z}}_1'|^2 \geq tn(1+\eta)).
    \end{split}
\end{equation}
Note that $\textbf{Q}\bar{\textbf{z}}_1'\sim \mathcal{N}(\textbf{0},\sigma^2\textbf{I}_{tn})$. Thus, $|\textbf{Q}\bar{\textbf{z}}_1'/\sigma|^2\sim \chi^2(tn)$. We can use the Chernoff bounding approach in \cite{Gamal04} to upper-bound $Pr(|\textbf{Q}\bar{\textbf{z}}_1'|^2 > tn(1+\eta))$, which gives
\begin{equation}
    \label{eq_wrong_decode_error_chernoff_bound}
    Pr(|\textbf{Q}\bar{\textbf{z}}_1'|^2 \geq tn(1+\eta)) \leq \min_{\alpha > 0}\; e^{-\alpha tn(1+\eta)} \mathbb{E}[e^{\alpha|\textbf{Q}\bar{\textbf{z}}_1'|^2}],
\end{equation}
where $\mathbb{E}[e^{\alpha|\textbf{Q}\bar{\textbf{z}}_1'|^2}] = \exp(-\frac{tn}{2}\ln(1-2\alpha \sigma^2))$. Therefore,
\begin{equation}
    \label{eq_wrong_decode_error_chernoff_bound_min}
    \begin{split}
        Pr(|\textbf{Q}\bar{\textbf{z}}_1'|^2 \geq tn(1+\eta)) & \leq \min_{\alpha > 0}\; e^{-\frac{tn}{2}[2\alpha (1+\eta)+\ln(1-2\alpha \sigma^2))]}\\
        & = \exp\left(-\frac{tn}{2}(\zeta - \ln \zeta - 1)\right),
    \end{split}
\end{equation}
where $\zeta = \frac{1+\eta}{\sigma^2}$. It is worth noting that $\sigma^2\rightarrow 1$ when $n\rightarrow\infty$. For arbitrary $\eta >0$, we have $\zeta > 1$ which leads to $\zeta - \ln \zeta - 1 >0$. We can then conclude that $Pr(\bar{\textbf{z}}_1 \notin \mathcal{E}_{n,\eta})\leq \epsilon_2/2$ for arbitrary $\eta, \epsilon_2>0$ and sufficiently large $n$. Finally, we complete the proof of (\ref{eq_achievable_rate_general}).

Since $\boldsymbol{\Sigma}_1(\textbf{W})$ can be further given by
\begin{equation}
    \label{eq_Sigma_1}
    \begin{split}
        \boldsymbol{\Sigma}_1(\textbf{W}) = & \sum_{l}\tilde a_l^2\textbf{I}_{tn} +  \textbf{W}\left(\textbf{I}_{rn} + \sum_{l} \bar{\textbf{H}}_l\bar{\textbf{K}}_l\bar{\textbf{H}}_l^T\right)\textbf{W}^T \\
        & - \left(\sum_{l}\tilde a_l \bar{\textbf{B}}_l^T\bar{\textbf{H}}_l^T\right) \textbf{W}^T - \textbf{W} \left(\sum_{l}\tilde a_l \bar{\textbf{H}}_l \bar{\textbf{B}}_l\right),
    \end{split}
\end{equation}
we can minimize $|\boldsymbol{\Sigma}_1(\textbf{W})|$ over $\textbf{W}$ by ``completing the square'', which gives
\begin{equation}
    \label{eq_optimal_W}
    \textbf{W}^* = \left(\sum_{l}\tilde a_l \bar{\textbf{B}}_l^T \bar{\textbf{H}}_l^T\right)\left(\textbf{I}_{rn} + \sum_{l} \bar{\textbf{H}}_l \bar{\textbf{K}}_l \bar{\textbf{H}}_l^T\right)^{-1}.
\end{equation}
Thus, 
\begin{equation}
    \label{eq_optimal_Sigma_1}
    \begin{split}
        \boldsymbol{\Sigma}_1(\textbf{W}^*) = &\sum_{l}\tilde a_l^2\textbf{I}_{tn} - \left(\sum_{l}\tilde a_l \bar{\textbf{B}}_l^T \bar{\textbf{H}}_l^T\right)\\
        &\cdot\left(\textbf{I}_{rn} + \sum_{l} \bar{\textbf{H}}_l \bar{\textbf{K}}_l \bar{\textbf{H}}_l^T\right)^{-1} \left(\sum_{l}\tilde a_l \bar{\textbf{H}}_l \bar{\textbf{B}}_l \right).
    \end{split}
    \end{equation}
To calculate $|\boldsymbol{\Sigma}_1(\textbf{W}^*)|$, we will use the matrix determinant lemma 
\begin{equation}\label{eq_matrix_determinant_lemma}
    |\textbf{D}-\textbf{C}\textbf{A}^{-1}\textbf{B}| = \frac{\left|\begin{array}{cc}
        \textbf{A} & \textbf{B} \\
        \textbf{C} & \textbf{D}
    \end{array}\right|}{|\textbf{A}|} = \frac{|\textbf{D}||\textbf{A}-\textbf{B}\textbf{D}^{-1}\textbf{C}|}{|\textbf{A}|},
\end{equation}
and Sylvester's determinant theorem: For a $m\times n$ matrix $\textbf{A}$ and a $n\times m$ matrix $\textbf{B}$, it holds that
\begin{equation}\label{eq_Sylvesters_determinant_theorem}
    |\textbf{I}_m+\textbf{A}\textbf{B}| = |\textbf{I}_n+\textbf{B}\textbf{A}|.
\end{equation}
We can simplify $|\boldsymbol{\Sigma}_1(\textbf{W}^*)|$ as
\begin{equation}
    \label{eq_optimal_det_Sigma_1}
    |\boldsymbol{\Sigma}_1(\textbf{W}^*)| = \frac{\left|\bar{\textbf{M}} \right|}{\left|\textbf{I}_{rn} + \sum_{l} \bar{\textbf{H}}_l \bar{\textbf{K}}_l\bar{\textbf{H}}_l^T \right|}, 
\end{equation}
where 
\begin{equation}
    \bar{\textbf{M}} = (\tilde a_1^2 + \tilde a_2^2)\textbf{I}_{tn} + (\tilde a_1 \bar{\textbf{B}}_2^T \bar{\textbf{H}}_2^T - \tilde a_2 \bar{\textbf{B}}_1^T \bar{\textbf{H}}_1^T)(\tilde a_1 \bar{\textbf{H}}_2 \bar{\textbf{B}}_2 - \tilde a_2 \bar{\textbf{H}}_1 \bar{\textbf{B}}_1).
\end{equation}
Therefore, the achievable rate to decode the first linear combination with coefficients $\textbf{a} = (a_1,a_2)^T$ is given by
\begin{equation}
    r_{l}^{(1)} = \frac{1}{2n} \log^+ 
    \frac{\beta_l^{2tn} \left|\textbf{I}_{rn} + \sum_{l} \bar{\textbf{H}}_l \bar{\textbf{K}}_l\bar{\textbf{H}}_l^T\right|}
    {\left|\bar{\textbf{M}}\right|}.
\end{equation}
Since $\textbf{I}_{rn} + \sum_{l} \bar{\textbf{H}}_l \bar{\textbf{K}}_l\bar{\textbf{H}}_l^T$ and $\bar{\textbf{M}}$ are both block diagonal matrix, $r_{l}^{(1)}$ can be written as
\begin{equation}
    \label{eq_achievable_rate_MIMO_1_sim}
    r_{l}^{(1)} = \frac{1}{2} \log^+ 
    \frac{\beta_l^{2t} \left|\textbf{I}_{r} + \sum_{l} \textbf{H}_l \textbf{K}_l\textbf{H}_l^T\right|}
    {\left|\textbf{M}\right|},
\end{equation}
where
\begin{equation}
    \textbf{M} = (\tilde a_1^2 + \tilde a_2^2)\textbf{I}_{t} + (\tilde a_1 \textbf{B}_2^T \textbf{H}_2^T - \tilde a_2 \textbf{B}_1^T \textbf{H}_1^T)(\tilde a_1 \textbf{H}_2 \textbf{B}_2 - \tilde a_2 \textbf{H}_1 \textbf{B}_1).
\end{equation}

Assume the first linear combination is decoded successfully, we can reconstruct $\sum_{l} \tilde a_l \textbf{c}_l = \sum_{l} a_l \tilde{\textbf{t}}_l + \sum_{l} a_l \beta_l \textbf{d}_l$. Let $\textbf{b} = (b_1,b_2)^T$ denote the computation coefficients of the second linear combination, such that the matrix $(\textbf{a}, \textbf{b})$ has full rank. The receiver computes
\begin{align*}
    \textbf{y}_2^\prime = & \textbf{F}\textbf{y} + \textbf{L} \sum_{l} a_l\beta_l \textbf{c}_l - \sum_{l} b_l \beta_l \textbf{d}_l \\
    = & \sum_l \textbf{F}\bar{\textbf{H}}_l \bar{\textbf{B}}_l \textbf{c}_l + \textbf{F}\textbf{z} + \textbf{L} \sum_{l} a_l\beta_l \textbf{c}_l - \sum_l b_l\beta_l \textbf{d}_l\\
    & - \sum_l b_l\beta_l \textbf{c}_l + \sum_l b_l\beta_l \textbf{c}_l \\
    = & \underbrace{\textbf{F}\bar{\textbf{z}} + \sum_l (\tilde b_l \textbf{I}_{tn} - \textbf{F}\bar{\textbf{H}}_l \bar{\textbf{B}}_l - \tilde a_l \textbf{L})(-\textbf{c}_l) }_{\bar{\textbf{z}}_2} + \sum_l b_l \tilde{\textbf{t}}.
\end{align*}
Using the same lattice decoding as the first linear combination, the achievable rate to decode $\sum_l b_l \tilde{\textbf{t}}_l$ is given by
\begin{equation}
    r_{l}^{(2)} = \max_{\textbf{F},\textbf{L}}\quad \frac{1}{2n} \log^+ \frac{|\beta_l^2\textbf{I}_{tn}|}{|\boldsymbol{\Sigma}_2(\textbf{F},\textbf{L})|}.
\end{equation}
The covariance matrix of the effective noise $\bar{\textbf{z}}_2$ is given by
\begin{equation}
    \label{eq_Sigma_2}
    \begin{split}
    \boldsymbol{\Sigma}_2(\textbf{F},\textbf{L}) = & \sum_{l}\tilde b_l^2 \textbf{I}_{tn} +  \textbf{F}\left(\textbf{I}_{rn} + \sum_{l} \bar{\textbf{H}}_l\bar{\textbf{K}}_l\bar{\textbf{H}}_l^T\right)\textbf{F}^T \\
    & - \left(\sum_{l}\tilde b_l \bar{\textbf{B}}_l^T\bar{\textbf{H}}_l^T\right) \textbf{F}^T - \textbf{F} \left(\sum_{l}\tilde b_l \bar{\textbf{H}}_l \bar{\textbf{B}}_l\right)\\
    & + \sum_l \tilde a_l^2\textbf{L}\textbf{L}^T - \left(\sum_l \tilde a_l (\tilde b_l \textbf{I}_{tn} - \textbf{F}\bar{\textbf{H}}_l \bar{\textbf{B}}_l)\right) \textbf{L}^T \\
    & - \textbf{L} \left(\sum_l \tilde a_l  (\tilde b_l \textbf{I}_{tn} - \bar{\textbf{B}}_l^T  \bar{\textbf{H}}_l^T \textbf{F}^T)\right)
    \end{split}
\end{equation}
Optimizing over $\textbf{L}$ gives
\begin{equation}
    \label{eq_optimal_L}
    \textbf{L}^* = \left(\sum_{l}\tilde a_l^2\right)^{-1}\left(\sum_l \tilde a_l (\tilde b_l \textbf{I}_{tn} - \textbf{F}\bar{\textbf{H}}_l \bar{\textbf{B}}_l)\right).
\end{equation}
Plugging this into (\ref{eq_Sigma_2}) gives
\begin{align*}
    \boldsymbol{\Sigma}_2(\textbf{F},\textbf{L}^*) = & \sum_{l}\tilde b_l^2 \textbf{I}_{tn} +  \textbf{F}\left(\textbf{I}_{rn} + \sum_{l} \bar{\textbf{H}}_l\bar{\textbf{K}}_l\bar{\textbf{H}}_l^T\right)\textbf{F}^T\\
    & - \left(\sum_{l}\tilde b_l \bar{\textbf{B}}_l^T\bar{\textbf{H}}_l^T\right) \textbf{F}^T - \textbf{F} \left(\sum_{l}\tilde b_l \bar{\textbf{H}}_l \bar{\textbf{B}}_l\right) \\    
    & - \left(\sum_{l}\tilde a_l^2\right)^{-1}\left(\sum_l \tilde a_l (\tilde b_l \textbf{I}_{tn} - \textbf{F}\bar{\textbf{H}}_l \bar{\textbf{B}}_l)\right) \\
    &\cdot\left(\sum_l \tilde a_l  (\tilde b_l \textbf{I}_{tn} - \bar{\textbf{B}}_l^T  \bar{\textbf{H}}_l^T \textbf{F}^T)\right)
\end{align*}
Simplifying $\boldsymbol{\Sigma}_2(\textbf{F},\textbf{L}^*)$ results in
\begin{equation}
    \label{eq_sigma_2_optimal_L}
    \begin{split} 
    \boldsymbol{\Sigma}_2(\textbf{F},\textbf{L}^*) = & \frac{(\tilde a_1 \tilde b_2 - \tilde a_2 \tilde b_1)^2}{\sum_{l}\tilde a_l^2}\textbf{I}_{tn} + \textbf{F} \textbf{M}_k \textbf{F}^T \\
    & - \frac{\tilde a_1 \tilde b_2 - \tilde a_2 \tilde b_1}{\sum_{l}\tilde a_l^2}
    [(\tilde a_1 \bar{\textbf{B}}_2^T\bar{\textbf{H}}_2^T - \tilde a_2 \bar{\textbf{B}}_1^T\bar{\textbf{H}}_1^T) \textbf{F}^T \\
    & + \textbf{F} (\tilde a_1 \bar{\textbf{H}}_2\bar{\textbf{B}}_2 - \tilde a_2 \bar{\textbf{H}}_1 \bar{\textbf{B}}_1)],
    \end{split}
\end{equation}
where 
\begin{equation}
    \label{eq_Mk}
    \begin{split}
        \textbf{M}_k = & \textbf{I}_{rn} + \left(\sum_{l}\tilde a_l^2\right)^{-1}(\tilde a_1 \bar{\textbf{H}}_2 \bar{\textbf{B}}_2 - \tilde a_2 \bar{\textbf{H}}_1 \bar{\textbf{B}}_1) \\
        & \cdot (\tilde a_1 \bar{\textbf{B}}_2^T \bar{\textbf{H}}_2^T - \tilde a_2 \bar{\textbf{B}}_1^T \bar{\textbf{H}}_1^T).
    \end{split}
\end{equation}
optimizing over $\textbf{F}$ gives
\begin{equation}
    \label{eq_optimal_F}
    \textbf{F}^* = \frac{\tilde a_1 \tilde b_2 - \tilde a_2 \tilde b_1}{\sum_{l}\tilde a_l^2}
    (\tilde a_1 \bar{\textbf{B}}_2^T\bar{\textbf{H}}_2^T - \tilde a_2 \bar{\textbf{B}}_1^T\bar{\textbf{H}}_1^T)
    \textbf{M}_k^{-1}.
\end{equation}
Thus,
\begin{equation}
    \label{eq_optimal_Sigma_2}
    \begin{split}
        \boldsymbol{\Sigma}_2(\textbf{F}^*,\textbf{L}^*) = \frac{(\tilde a_1 \tilde b_2 - \tilde a_2 \tilde b_1)^2}{\sum_{l}\tilde a_l^2}\textbf{I}_{tn} - \left(\frac{\tilde a_1 \tilde b_2 - \tilde a_2 \tilde b_1}{\sum_{l}\tilde a_l^2} \right)^2 \\
        \cdot (\tilde a_1 \bar{\textbf{B}}_2^T\bar{\textbf{H}}_2^T - \tilde a_2 \bar{\textbf{B}}_1^T\bar{\textbf{H}}_1^T) \textbf{M}_k^{-1} (\tilde a_1 \bar{\textbf{H}}_2\bar{\textbf{B}}_2 - \tilde a_2 \bar{\textbf{H}}_1 \bar{\textbf{B}}_1).
    \end{split}
\end{equation}
To calculate the determinant, we apply the matrix determinant lemma (\ref{eq_matrix_determinant_lemma}), which leads to
\begin{align}
    |\boldsymbol{\Sigma}_2(\textbf{F}^*,\textbf{L}^*)|
    = & \frac{(\tilde a_1 \tilde b_2 - \tilde a_2 \tilde b_1)^{2tn}}
    {\left|\bar{\textbf{M}}\right|} \label{eq_optimal_det_Sigma_2}
\end{align}
Therefore, given the first linear combination with coefficients $\textbf{a} = (a_1,a_2)^T$, the achievable rate to decode the second linear combination with coefficients $\textbf{b} = (b_1,b_2)^T$ is
\begin{equation}
    r_{l}^{(2)} = \frac{1}{2n} \log^+ 
    \frac{\beta_l^{2tn} {\left|\Bar{\textbf{M}}\right|}}
    {(\tilde a_1 \tilde b_2 - \tilde a_2 \tilde b_1)^{2tn}}.
\end{equation}
Since $\bar{\textbf{M}}$ is block diagonal, $r_{l}^{(2)}$ can be written as
\begin{equation}
    \label{eq_achievable_rate_MIMO_2_sim}
    r_{l}^{(2)} = \frac{1}{2} \log^+ 
    \frac{\beta_l^{2t} {\left|\textbf{M}\right|}}
    {(\tilde a_1 \tilde b_2 - \tilde a_2 \tilde b_1)^{2t}}.
\end{equation}
To have both linear combinations decoded successfully, the transmission rate of each user should not exceed the smaller achievable rate of both decoding processes. If $a_l$ or $b_l$ equals zero, which means one linear combination contains no information of user $l$. The achievable rate of user $l$ simply comes from the other linear combination. Therefore, the proof of Theorem~\ref{thm_rate_pair} is complete.
\end{IEEEproof}

\begin{remark}
    when $t=r=1$, the results of Theorem~\ref{thm_rate_pair} reduce to Theorem~$2$ in \cite{Zhu17}. 
\end{remark}

The sum capacity of a two-user MIMO Gaussian MAC with channel matrices $\textbf{H}_1,\textbf{H}_2$ is known to be \cite{gamal_kim_2011}
\begin{equation}
    \label{eq_sum_capacity}
    C_{sum} = \max_{\textbf{K}_1,\textbf{K}_2} \;\frac{1}{2}\log \left|\textbf{I}_{r} + \textbf{H}_1 \textbf{K}_1\textbf{H}_1^T + \textbf{H}_2 \textbf{K}_2\textbf{H}_2^T\right|,
\end{equation}
where the optimized input covariance matrices $\textbf{K}_1^*, \textbf{K}_2^*$ can be found by the iterative water-fill algorithm \cite{gamal_kim_2011}. Our derived results in Theorem~\ref{thm_rate_pair} hold for any input covariance matrices hence including the optimal $\textbf{K}_1$ and $\textbf{K}_2$. To simplify the presentation in the rest of this paper, for given $\textbf{H}_1$ and $\textbf{H}_2$ we define 
\begin{equation}
    \label{eq_c_sum_wo_log}
    C_d = \left|\textbf{I}_{r} + \textbf{H}_1 \textbf{K}_1^*\textbf{H}_1^T + \textbf{H}_2 \textbf{K}_2^*\textbf{H}_2^T\right|.
\end{equation}

\begin{lem}\label{lem_sum_capacity}
    For a two-user MIMO Gaussian MAC given the channel matrices $\textbf{H}_1$ and $\textbf{H}_2$, with the CFMA scheme, the sum capacity is achievable by choosing $\textbf{a} = (1,1)$, $\textbf{b} = (1,0)$ or $(0,1)$, and $\beta_1/\beta_2 = \gamma$, if there exits some real positive $\gamma$ such that
    \begin{equation}
    \label{ineq_beta_condition_sumrate}
    f(\gamma) - \gamma^{t} \sqrt{C_d} \leq 0,
\end{equation}
where 
\begin{equation}\label{eq_f_gamma}
    f(\gamma) = \left|(\gamma^2+1)\textbf{I}_{t} + (\gamma \textbf{B}_2^{*T} \textbf{H}_2^T - \textbf{B}_1^{*T} \textbf{H}_1^T)(\gamma \textbf{H}_2 \textbf{B}_2^* - \textbf{H}_1 \textbf{B}_1^*)\right|.
\end{equation}
In particular, $\textbf{B}_l^{*}$ are $t\times t$ lower triangular matrices such that $\textbf{B}_l^{*}\textbf{B}_l^{*T} = \textbf{K}_l^*$ for $l=1,2$.
\end{lem}
\begin{IEEEproof}
    When we choose $\textbf{a} = (1,1)$ and $\textbf{b} = (1,0)$, the achievable rate of user $2$ is given by $R_2 = r_2(\textbf{a},\boldsymbol{\beta})$ according to Theorem~\ref{thm_rate_pair}. If (\ref{ineq_beta_condition_sumrate}) holds, with $\gamma = \beta_1/\beta_2$ we have $\left| \textbf{M} \right|^2\leq \beta_1^{2t}\beta_2^{2t}C_d$, where $\textbf{M}$ is given in Theorem~\ref{thm_rate_pair} with $\textbf{B}_l = \textbf{B}_l^{*}$. It can be further inferred that $r_1(\textbf{b}|\textbf{a},\boldsymbol{\beta})\leq r_1(\textbf{a},\boldsymbol{\beta})$. Thus the achievable rate of user $1$ is given by $R_1 = r_1(\textbf{b}|\textbf{a},\boldsymbol{\beta})$. The achievable sum rate in this case can be written as
    \begin{equation}
        \label{eq_sum_rate_b10}
        R_{sum} = R_1+R_2 = C_{sum} - t\log |a_1b_2-a_2b_1|.
    \end{equation}
    Since $|a_1b_2-a_2b_1| = 1$ for the chosen $\textbf{a}$ and $\textbf{b}$, the sum rate achieves the sum capacity. Similarly, when $\textbf{a} = (1,1)$ and $\textbf{b} = (0,1)$, we have $R_1 = r_1(\textbf{a},\boldsymbol{\beta})$ and $R_2 = r_2(\textbf{b}|\textbf{a},\boldsymbol{\beta})$. The achievable sum rate is also given by (\ref{eq_sum_rate_b10}), thus the proof is complete.
\end{IEEEproof}

\begin{remark}
    It should be pointed out that for given channel coefficients and power constraints, (\ref{ineq_beta_condition_sumrate}) may not be satisfied for any choice of $\gamma$ as we will show in Section~\ref{subsec_sim} with some numerical examples. Thus, the sum capacity of the Gaussian MIMO channel is not always achievable with the current CFMA scheme. This is in contrast to the result \cite{Zhu17} in the SISO case, where it is shown that the sum capacity (in fact any rate pair on the entire dominant face) is always achievable when the signal-to-noise ratio is high enough. But also notice that this result is derived from a specific code construction, where a long lattice codebook (with length $tn$) is constructed described in Section~\ref{sec_coding}, and the transmitted codeword is "chopped" into $t$ parts and each piece is fed into one antenna for transmission (after precoding). It remains unclear if this scheme can be improved and if a further modified CFMA-type coding scheme is capacity-achieving for all channel coefficients. 
\end{remark}

\begin{remark}
    For given channel states and channel input covariance matrices, the left-hand side (LHS) of (\ref{ineq_beta_condition_sumrate}), denoted as $g(\gamma)=f(\gamma) - \gamma^t \sqrt{C_d}$, is a polynomial of $\gamma$ with the highest order $2t$. Its leading coefficient is given by $|\textbf{I}_t+\textbf{B}_2^{*T} \textbf{H}_2^{T}\textbf{H}_2 \textbf{B}_2^*|$, which is always positive. Thus, (\ref{ineq_beta_condition_sumrate}) can be satisfied if and only if $g(\gamma)$ has real roots. There are several numerical ways to check if a polynomial has real roots, for example, Sturm's theorem \cite{Basu16AlgorithmAlgebra}.
\end{remark}

\section{Case study}\label{sec_case}
Apart from the numerical methods to check if the polynomial of interest has real roots, we study some special cases and provide analytical capacity-achieving conditions for them.

\subsection{SIMO MAC}
We consider the special case when each transmitter only has a single transmit antenna, i.e., $t = 1$, while the receiver has $r\geq1$ antennas. The channel coefficients are now given by vectors $\textbf{h}_1$ and $\textbf{h}_2$ of length $r$, and $K_1 = K_2 = P$ are two scalars. In this case, $C_d$ in (\ref{eq_c_sum_wo_log}) is given by
\begin{equation}
    \label{eq_sumcapacity_simo}
    C_d = \left|\textbf{I}_{r} + (\textbf{h}_1 \textbf{h}_1^T + \textbf{h}_2 \textbf{h}_2^T) P\right|.
\end{equation}
Note that the matrix $\textbf{h}_1 \textbf{h}_1^T + \textbf{h}_2 \textbf{h}_2^T$ is semi-positive definite and can be decomposed as $\textbf{U} \boldsymbol{\Lambda} \textbf{U}^T$, where $\textbf{U}$ is unitary, i.e., $\textbf{U}^T\textbf{U} = \textbf{I}_r$ and $\boldsymbol{\Lambda}$ is diagonal whose diagonal entries are the eigenvalues of $\textbf{h}_1 \textbf{h}_1^T + \textbf{h}_2 \textbf{h}_2^T$. Due to its special structure, the rank of $\textbf{h}_1 \textbf{h}_1^T + \textbf{h}_2 \textbf{h}_2^T$ is smaller than or equal to $2$. So it has at most two positive eigenvalues defined as $\lambda_1, \lambda_2$, and the other eigenvalues (if they exist) will be zero. Specifically, when $\textbf{h}_1$ and $\textbf{h}_2$ are collinear, we have $\lambda_2 = 0$ and $\lambda_1 = ||\textbf{h}_1||^2+||\textbf{h}_2||^2$ by calculating the roots of the characteristic polynomial of $\textbf{h}_1 \textbf{h}_1^T + \textbf{h}_2 \textbf{h}_2^T$ \cite{Andrilli23LinearAlgebra}. Thus, from Sylvester’s determinant theorem \cite{Sylvester1851}, (\ref{eq_sumcapacity_simo}) can be written as
\begin{equation}\label{eq_sumcapacity_simo_eigen_collinear}
    C_d = \left|\textbf{I}_{r} +  P \boldsymbol{\Lambda}\right| = 1+\lambda_1 P.
\end{equation}
If $\textbf{h}_1$ and $\textbf{h}_2$ are not collinear, we have $\lambda_1, \lambda_2>0$. In this case,  
\begin{equation}
    \label{eq_sumcapacity_simo_eigen}
    C_d = (1+\lambda_1 P)(1+\lambda_2 P),
\end{equation}
% where $\lambda_1+\lambda_2 = ||\textbf{h}_1||^2+||\textbf{h}_2||^2$ since $\lambda_1+\lambda_2 = tr(\boldsymbol{\Lambda}) = tr(\textbf{h}_1 \textbf{h}_1^T + \textbf{h}_2 \textbf{h}_2^T) = tr(\textbf{h}_1 \textbf{h}_1^T) + tr(\textbf{h}_2 \textbf{h}_2^T)$ and $tr(\textbf{h}_l \textbf{h}_l^T) = tr(\textbf{h}_l^T\textbf{h}_l) = ||\textbf{h}_l||^2$ for $l = 1,2$. 
By analysing the capacity-achieving condition (\ref{ineq_beta_condition_sumrate}), we have the following lemma.
\begin{lem}\label{lem_SIMO}
    For a two-user SIMO Gaussian MAC given the channel coefficients $\textbf{h}_1$ and $\textbf{h}_2$, the sum capacity is achievable with CFMA if $\Delta \geq 0$, where
    \begin{equation}
        \label{eq_disc_simo}
        \Delta = (\sqrt{C_d} + 2 P \textbf{h}_1^T \textbf{h}_2)^2 - 4(1+P||\textbf{h}_1||^2)(1+P||\textbf{h}_2||^2).
    \end{equation}
    In particular, $\gamma$ should be chosen within
    \begin{equation}
        \label{range_gamma_simo}
        \left[\frac{\sqrt{C_d} + 2 P \textbf{h}_1^T \textbf{h}_2-\sqrt{\Delta}}{2(1+P||\textbf{h}_2||^2)} ,\frac{\sqrt{C_d} + 2 P \textbf{h}_1^T \textbf{h}_2 + \sqrt{\Delta}}{2(1+P||\textbf{h}_2||^2)}\right],
    \end{equation}
    where $C_d$ is given by (\ref{eq_sumcapacity_simo}).
\end{lem}
\begin{IEEEproof}
    When $t=1$, $f(\gamma)$ in Lemma~\ref{lem_sum_capacity} is given by
    \begin{equation}
        \label{eq_f_gamma_simo}
        f(\gamma) = (\gamma^2+1) + (\gamma \sqrt{P} \textbf{h}_2^T - \sqrt{P} \textbf{h}_1^T)(\gamma \textbf{h}_2 \sqrt{P} - \textbf{h}_1 \sqrt{P}).
    \end{equation}
    The LHS of (\ref{ineq_beta_condition_sumrate}) in Lemma~\ref{lem_sum_capacity} is given by
    \begin{equation}
        \label{eq_g_gamma_simo}
        g(\gamma) = (1+P||\textbf{h}_2||^2)\gamma^2 - (\sqrt{C_d} + 2 P \textbf{h}_1^T \textbf{h}_2)\gamma + (1+P||\textbf{h}_1||^2).
    \end{equation}
    Note that $g(\gamma)$ is a second-order polynomial of $\gamma$ with positive leading coefficient. Its discriminant $\Delta$ is given by (\ref{eq_disc_simo}). When $\Delta \geq 0$, $g(\gamma)$ will have two real roots and (\ref{ineq_beta_condition_sumrate}) is satisfied if $\gamma$ is within the two roots. Notice that both roots given by (\ref{range_gamma_simo}) are positive since $\sqrt{C_d} + 2 P \textbf{h}_1^T \textbf{h}_2>0$. Otherwise, if $\sqrt{C_d} + 2 P \textbf{h}_1^T \textbf{h}_2\leq 0$, we have $(\sqrt{C_d} + 2 P \textbf{h}_1^T \textbf{h}_2)^2 < (2 P \textbf{h}_1^T \textbf{h}_2)^2 \leq 4P^2||\textbf{h}_1||^2||\textbf{h}_2||^2$. Thus, $\Delta$ will become negative, which conflicts with $\Delta \geq 0$. Therefore, the sum capacity is achievable in this case and the proof is complete.
\end{IEEEproof}

\begin{remark}
    If we further let $r=1$, the channel becomes the SISO MAC. The sum capacity-achieving condition in Lemma~\ref{lem_SIMO} specializes to Theorem 3 Case II) in  \cite{Zhu17}.
\end{remark}

The following corollary shows that it is possible to achieve the sum capacity when the transmission power is large enough with some channel conditions.
\begin{coro}\label{coro_simo}
    Consider a two-user SIMO Gaussian MAC given the channel coefficients $\textbf{h}_1$ and $\textbf{h}_2$.
    \begin{enumerate}
        \item When $\textbf{h}_1$ and $\textbf{h}_2$ are collinear, the sum capacity is achievable with CFMA if 
        \begin{equation}\label{ineq_power_channel_condtion_simo_col}
            \frac{P \textbf{h}_1^T \textbf{h}_2}{\sqrt{1+P(||\textbf{h}_1||^2+||\textbf{h}_2||^2)}} \geq \frac{3}{4}
        \end{equation}
        and $\gamma$ is chosen within (\ref{range_gamma_simo}).
        \item When $\textbf{h}_1$ and $\textbf{h}_2$ are linearly independent, the sum capacity is achievable with CFMA if 
        \begin{equation}\label{ineq_channel_condtion_simo_noncol}
            (\sqrt{\lambda_1\lambda_2} + 2\textbf{h}_1^T \textbf{h}_2)^2 > 4||\textbf{h}_1||^2||\textbf{h}_2||^2,
        \end{equation}
        and if $P\geq P^*$, $\gamma$ is chosen within (\ref{range_gamma_simo}), where $P^*$ is the largest root of $\Delta(P)$ and $\Delta(P)$ is given by (\ref{eq_disc_simo}) where we view $\Delta$ as a polynomial of $P$.
    \end{enumerate}
\end{coro}
\begin{IEEEproof}
    When $\textbf{h}_1$ and $\textbf{h}_2$ are collinear, we have $(\textbf{h}_1^T \textbf{h}_2)^2 = ||\textbf{h}_1||^2||\textbf{h}_2||^2$ and $C_d = 1+P(||\textbf{h}_1||^2+||\textbf{h}_2||^2)$ given by (\ref{eq_sumcapacity_simo_eigen_collinear}). Therefore, $\Delta$ in (\ref{eq_disc_simo}) can be simplified as
    \begin{equation}\label{eq_disc_simo_col}
        \Delta = 4 P \textbf{h}_1^T \textbf{h}_2 \sqrt{C_d} - 3C_d.
    \end{equation}
    If (\ref{ineq_power_channel_condtion_simo_col}) holds, we have $\Delta\geq 0$. Thus, the sum capacity is achievable in this case according to Lemma~\ref{lem_SIMO}.
    
    When $\textbf{h}_1$ and $\textbf{h}_2$ are linearly independent, if (\ref{ineq_channel_condtion_simo_noncol}) holds and with (\ref{eq_sumcapacity_simo_eigen}), we have $\Delta(P)$ given by (\ref{eq_disc_simo}) becoming a 2nd-order polynomial of $P$ with positive leading coefficient. Thus, as $P$ becomes large enough, $\Delta(P)$ will eventually become non-negative. Since $\Delta(0) = -3$, the largest root of $\Delta(P)$, denoted by $P^*$, will always be positive. If $P\geq P^*$, we have $\Delta(P)\geq 0$. Thus, the sum capacity is achievable in this case according to Lemma~\ref{lem_SIMO}.
\end{IEEEproof}

\subsection{Diagonal $2$-by-$2$ MIMO MAC}
We consider the special case when $\textbf{H}_1$ and $\textbf{H}_2$ are diagonal. In this case, the diagonal input covariance matrices with optimized power splitting can achieve the sum capacity. When $t=r=2$, we can denote
\begin{equation}
\label{eq_h_k_diagonal}
    \textbf{H}_l = \left[\begin{array}{cc}
        h_{l1} &  \\
         & h_{l2}
    \end{array}\right],
    \textbf{K}_l^* = \left[\begin{array}{cc}
        k_{l1} &  \\
         & k_{l2}
    \end{array}\right],
\end{equation}
where $k_{l1}, k_{l2}$ are the optimal power splitting and $k_{l1} + k_{l2} = P$ for $l=1,2$. We define some auxiliary parameters $c_{l1},c_{l2}$ for $l = 1,2$ as
\begin{equation}
    \label{eq_c_lj}
    c_{l1} = h_{l1}\sqrt{\frac{k_{l1}}{P}},\quad c_{l2} = h_{l2}\sqrt{\frac{k_{l2}}{P}}.
\end{equation}
For the given finite $\textbf{H}_1$ and $\textbf{H}_2$, $c_{l1}$ and $c_{l2}$ always take finite values since they only contain the channel coefficient and power splitting ratio. We give a sufficient condition that guarantees to achieve the sum capacity for this case in the following lemma.
\begin{lem}\label{lem_diagonal}
    For a $2$-by-$2$ diagonal MIMO Gaussian MAC with channel matrices $\textbf{H}_1$ and $\textbf{H}_2$, when the power $P$ is large enough, the sum capacity is achievable with CFMA if either of the following conditions is satisfied.
    \begin{enumerate}
        \item $\gamma = c_{11}/c_{21}$ and
        \begin{equation}
            \label{ineq_condition_diagonal_1}
            \left(\frac{c_{22}}{c_{21}} - \frac{c_{12}}{c_{11}}\right)^2 < \sqrt{\frac{c_{12}^2+c_{22}^2}{c_{11}^2+c_{21}^2}};
        \end{equation}
        \item $\gamma = c_{12}/c_{22}$ and
        \begin{equation}
            \label{ineq_condition_diagonal_2}
            \left(\frac{c_{21}}{c_{22}} - \frac{c_{11}}{c_{12}}\right)^2 < \sqrt{\frac{c_{11}^2+c_{21}^2}{c_{12}^2+c_{22}^2}}.
        \end{equation}
    \end{enumerate}
\end{lem}
\begin{IEEEproof}
    With $\textbf{H}_l$ and $\textbf{K}_l^*$ given by (\ref{eq_h_k_diagonal}), $f(\gamma)$ in (\ref{ineq_beta_condition_sumrate}) is given by
    \begin{equation}
        \label{eq_f_gamma_diagonal}
        \begin{split}
            f(\gamma)
            % = & \prod_{i=1}^2 \gamma^2+1+(\gamma \sqrt{k_{2i}}h_{2i}-\sqrt{k_{1i}}h_{1i})^2 \\
            = & \prod_{i=1}^2 \gamma^2+1+(\gamma c_{2i}-c_{1i})^2 P,
        \end{split}
    \end{equation}
    and (\ref{eq_c_sum_wo_log}) is given by
    \begin{equation}
        \label{eq_sum_capacity_diagonal}
        C_d = \prod_{i=1}^2 1+h_{1i}^2k_{1i} + h_{2i}^2k_{2i} = \prod_{i=1}^2 1+(c_{1i}^2 + c_{2i}^2)P.
    \end{equation}
    The capacity-achieving condition (\ref{ineq_beta_condition_sumrate}) is equivalent to 
    \begin{equation}
        \label{eq_q_gamma}
        q(\gamma) = f(\gamma)^2 - \gamma^{2t} C_d \leq 0.
    \end{equation}
    Following conditions 1) in Lemma~\ref{lem_diagonal}, we have $\gamma c_{21} - c_{11} = 0$. With this condition, $f(\gamma)$ in (\ref{eq_f_gamma_diagonal}) can be written as 
    \begin{equation}
        \label{eq_f_gamma_diagonal_1}
        f(\gamma) = (\gamma^2+1)(\gamma^2+1+(\gamma c_{22}-c_{12})^2 P).
    \end{equation}
    Therefore, with this given $\gamma$, $q(\gamma)$ in (\ref{eq_q_gamma}) becomes a 2nd-order polynomial of $P$ with the leading coefficient given by 
    \begin{align*}
        & (\gamma^2+1)^2(\gamma c_{22}-c_{12})^4 - \gamma^4 (c_{11}^2 + c_{21}^2)(c_{12}^2 + c_{22}^2) \\
        = & \gamma^4 \left[(1+\frac{1}{\gamma^2})^2(\gamma c_{22}-c_{12})^4 - (c_{11}^2 + c_{21}^2)(c_{12}^2 + c_{22}^2)\right] \\
        = & \gamma^4 \left[(1+\frac{c_{21}^2}{c_{11}^2})^2(\frac{c_{11}c_{22}}{c_{21}}-c_{12})^4 - (c_{11}^2 + c_{21}^2)(c_{12}^2 + c_{22}^2)\right] \\
        = & \gamma^4 (c_{11}^2 + c_{21}^2)^2 \left[(\frac{c_{22}}{c_{21}}-\frac{c_{12}}{c_{11}})^4 - \frac{c_{12}^2 + c_{22}^2}{c_{11}^2 + c_{21}^2}\right]
    \end{align*}
    According to (\ref{ineq_condition_diagonal_1}), this leading coefficient is negative. Thus when $P$ is large enough, $q(\lambda)$ will become negative, which satisfies (\ref{ineq_beta_condition_sumrate}) in Lemma~\ref{lem_sum_capacity} and the sum capacity is achievable. Condition 2) in Lemma~\ref{lem_diagonal} is symmetric to Condition 1) and can be proved through the same procedure. 
\end{IEEEproof}
\begin{remark}
    It is worth noting that in both conditions of Lemma~\ref{lem_diagonal}, $\gamma$ is chosen as a constant. Therefore, $q(\gamma)$ in (\ref{eq_q_gamma}) can be regarded as a quadratic function of $P$ with a negative leading coefficient. Since $q(\gamma)>0$ when $P=0$, this quadratic function always has a positive real root. We can lower bound $P$ by this root to achieve the sum capacity. However, since the expression is complicated and due to the page limitation, we omit that lower bound in the lemma.
\end{remark}

\subsection{Simulation results}\label{subsec_sim}
To intuitively understand whether CFMA can achieve the sum capacity under different channel conditions, we conduct numerical studies with different channel models, including SIMO MAC, MAC with diagonal MIMO channels, and MAC with generic MIMO channels. For each realization, the channel coefficients are randomly generated from a uniform distribution over the interval $[0,1]$. We check whether CFMA can achieve the sum capacity by (\ref{ineq_beta_condition_sumrate}) with all real positive $\gamma$. For each case, we generate $10^4$ realizations and count the number when the sum capacity is achievable with CFMA (i.e., if (\ref{ineq_beta_condition_sumrate}) is satisfied) for different power constraints. Then we can calculate the ratio of this number to the total number of realizations. We will denote this ratio as $R_A$ in the later discussion. Among all the channel realizations, we observe that a large proportion, which is around $0.88$ of the channel realizations satisfy (\ref{ineq_channel_condtion_simo_noncol}) in the SIMO case. 

\begin{figure}
    \centering
    \includegraphics[width = 0.95\linewidth]{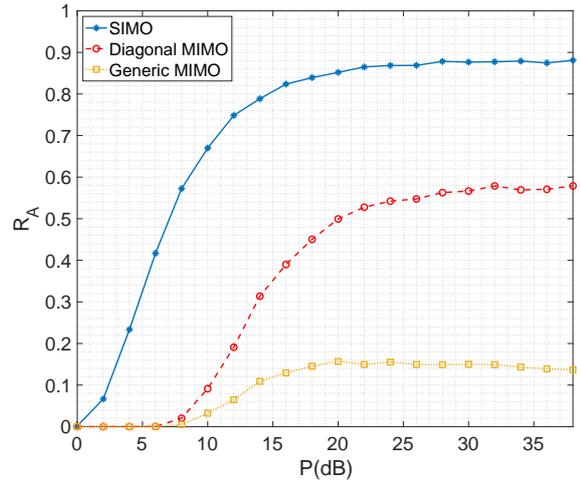}
    \caption{Simulation results for SIMO MAC and $2$-by-$2$ MIMO MAC}
    \label{fig_SIMO_MIMO}
\end{figure}

We plot $R_A$ against the power constraint $P$ in Fig.~\ref{fig_SIMO_MIMO} for the discussed channel models. The solid curve refers to the SIMO case with $r=2$. The dashed and dotted curves refer to the $2$-by-$2$ MIMO case with diagonal and generic channels, respectively. We observe that in the SIMO and diagonal MIMO cases, $R_A$ grows with $P$ and reaches a ceiling of around $0.88$ from $P=25$ dB and $0.58$ from $P=30$ dB, respectively. The value of $R_A$ in the SIMO case matches the proportion of the channel realizations which satisfy (\ref{ineq_channel_condtion_simo_noncol}), thus verifying Corollary~\ref{coro_simo}. In the generic $2$-by-$2$ MIMO case, it becomes much harder for CFMA to achieve the sum capacity. We can observe that $R_A$ is always below $0.2$ for different $P$. Moreover, the largest $R_A$ is achieved at $P=20$ dB, and then $R_A$ decreases as $P$ increases. This means for some specific channel conditions, a large power may not benefit CFMA to achieve the sum capacity (under the current code construction). 

% \section*{Acknowledgments}
% This work was supported in part by the Australian Research Council under Project DE210101497.

\section{Conclusion}
In this paper, we consider CFMA as the multiple-access technique for the two-user MIMO MAC. We characterize the achievable rate pair with CFMA for the general MIMO MAC with an arbitrary number of transmit and receive antennas. We show that it is possible to achieve the sum capacity given proper power and channel conditions. In future work, other variations of the CFMA scheme (for example, considering individual encoding for each antenna) will be investigated for their capacity-achieving capability. In addition, by continuing the successive computing for more linear combinations, it is possible to extend the results to multiple-user cases with more than two users.

\bibliographystyle{IEEEtran}
\bibliography{reference,IEEEabrv}

\end{document}